\documentclass[twocolumn,aps,prl,groupedaddress,nofootinbib,showpacs]{revtex4}%
\usepackage{graphicx}
\usepackage{amsmath}
\usepackage{bm}

\usepackage{relsize}
\RequirePackage{xspace}

\begin{document}


\title{\mbox{}\\[10pt]
QCD corrections to $\bm{e^+ e^- \to J/\psi g  g}$ at B Factories}

\author{Yan-Qing Ma$~^{(a)}$, Yu-Jie Zhang$~^{(a)}$, and Kuang-Ta Chao$~^{(a,b)}$}
\affiliation{ {\footnotesize (a)~Department of Physics and State Key
Laboratory of Nuclear Physics and Technology, Peking University,
 Beijing 100871, China}\\
{\footnotesize (b)~Center for High Energy Physics, Peking
University, Beijing 100871, China}}




\begin{abstract}
In heavy quarkonium production, the measured ratio $R_{c \bar
c}=\sigma[J/\psi+ c\bar c + X]/\sigma[J/\psi +X]$ at B factories is
much larger than existing theoretical predictions. To clarify this
discrepancy, in nonrelativistic QCD (NRQCD) we find the
next-to-leading-order (NLO) QCD correction to $e^+ e^- \to J/\psi
+gg$ can enhance the cross section by about 20\%. Together with the
calculated NLO result for $e^+ e^- \to J/\psi +c\bar c$, we show
that the NLO corrections can significantly improve the fit to the
ratio $R_{c \bar c}$. The effects of leading logarithm resummation
near the end point on the $J/\psi$ momentum distribution and total
cross section are also considered. Comparison of the calculated
cross section for $e^+ e^- \to J/\psi +gg$ with observed cross
section for $e^+ e^- \to J/\psi +non(c\bar c)$ is expected to
provide unique information on the issue of color-octet
contributions.
\end{abstract}

\maketitle


In recent years a number of challenging problems in heavy quarkonium
production have appeared\cite{Brambilla:2004wf}. Aside from the
$J/\psi$ production cross sections and polarizations in
hadron-collisions at the Tevatron,  charmonium production in
$e^+e^-$ annihilation at B factories\cite{Abe:2002rb,BaBar:2005}
also conflicted with theoretical predictions. The observed double
charmonium production cross sections for $e^+e^-\to
J/\psi\eta_c(\chi_{c0})$ were larger than the LO calculations in
NRQCD\cite{Bodwin:1994jh} by an order of
magnitude\cite{Braaten:2002fi}, and later it was found that these
discrepancies could be largely resolved by the NLO QCD corrections
(see \cite{Zhang:2005cha,Gong:2007db} for $J/\psi\eta_c$ and
\cite{zhangma08} for $J/\psi\chi_{c0}$) with relativistic
corrections\cite{bodwin06,He:2007te}. For the $J/\psi$ production
associated with an open charm pair $e^+ e^- \to J/\psi + c \bar c$,
the NLO QCD correction\cite{Zhang:2006ay} was also found to
significantly enhance the cross section (see also
\cite{Nayak:2007zb}), and reduce the large gap between experiment
and the LO calculations\cite{cs}.

Another important issue concerns the ratio
\begin{equation}
\label{Rdef}
 R_{c \bar c}=\frac{\sigma[e^+e^- \to J/\psi + c \bar c + X]}
{\sigma[e^+e^- \to J/\psi +X]}.
\end{equation}
Belle found first
$R_{c \bar c}=0.59 ^{+0.15}_{-0.13}\pm 0.12$\cite{Abe:2002rb}, and
later
 $R_{c \bar c} = 0.82 \pm 0.15 \pm 0.14$ \cite{Uglov:2004xa}.
On the contrary, LO NRQCD\cite{cs,Hagiwara:2004pf} and light-cone
pQCD predictions\cite{Berezhnoy:2003hz} for the ratio are only about
$0.1-0.3$.
The color evaporation model
gives a value of only 0.06\cite{Kang:2004zj}.

In NRQCD, $\sigma[J/\psi +X]$ includes color-singlet contributions
$\sigma[J/\psi(^3S_1^{[1]}) +c \bar c]$ and $\sigma[
J/\psi(^3S_1^{[1]}) +gg]$, and color-octet contributions
$\sigma[J/\psi(^3P_J^{[8]},^1S_0^{[8]}) +g]$. Contributions of other
Fock states are suppressed by either $\alpha_s$, the strong coupling
constant, or $v$, the relative velocity between quark and antiquark
in heavy quarkonium. $\sigma[J/\psi(^3P_J^{[8]},^1S_0^{[8]}) +g]$
was calculated at LO in $\alpha_s$\cite{Braaten:1995ez}, and an
apparent enhancement at the $J/\psi$ maximum energy was predicted.
But experiments did not show any enhancement at the end point. The
resummations of the $v$ expansion and $\log (1-z)$ where $z=E_{c\bar
c}/E_{c\bar c}^{max}$ are considered\cite{Fleming:2003gt}, but the
theoretical results rely heavily on the phenomenological shape
function.
It is possible that the observed end point behavior of $J/\psi$ and
the large ratio $R_{c \bar c}$ might indicate that the color-octet
matrix elements are much smaller than previously expected. To test
this thought we assume the color-octet contribution to be ignored
and only consider the color-singlet contributions. Under this
assumption, the ratio becomes
\begin{equation}
\label{RdefCS} R_{c \bar c}=\sigma[J/\psi+ c\bar c]/(\sigma[J/\psi+
c\bar c]+\sigma[J/\psi +gg]).
\end{equation}

In the following  we concentrate on $\sigma(J/\psi +gg)$ in NRQCD.
Aside from the LO calculations in NRQCD (see related references in
\cite{cs,Hagiwara:2004pf}), Ref.\cite{Lin:2004eu} considered
$\sigma[ J/\psi+ g g]$ within the framework of soft collinear
effective theory (SCET), and Ref.\cite{Leibovich:2007vr} summed over
the leading and next-to-leading logarithms in the end point region
of  $\sigma[ J/\psi+ g g]$. However, considering the crucially
importance of the NLO QCD corrections found in many heavy quarkonium
production
processes\cite{Zhang:2005cha,Gong:2007db,zhangma08,Zhang:2006ay,Nayak:2007zb,NLO},
it is necessary to carry out the calculation of NLO QCD correction
to $e^+ e^- \to J/\psi + g g$.

\begin{figure}
\includegraphics[width=8.5cm]{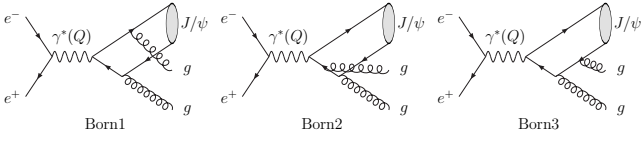}
\caption{\label{fig:LO}Three of the six Born diagrams for $e^+e^-
\to J/\psi gg$. }
\end{figure}

\begin{figure}
\includegraphics[width=8.5cm]{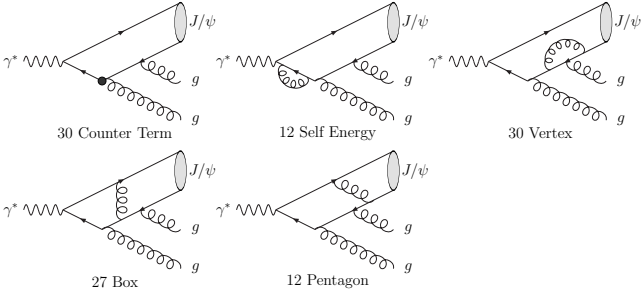}
\caption{\label{fig:vir} Feynman diagrams  for the virtual
correction to $e^+e^-  \to J/\psi gg$. }
\end{figure}

\begin{figure}
\includegraphics[width=8.5cm]{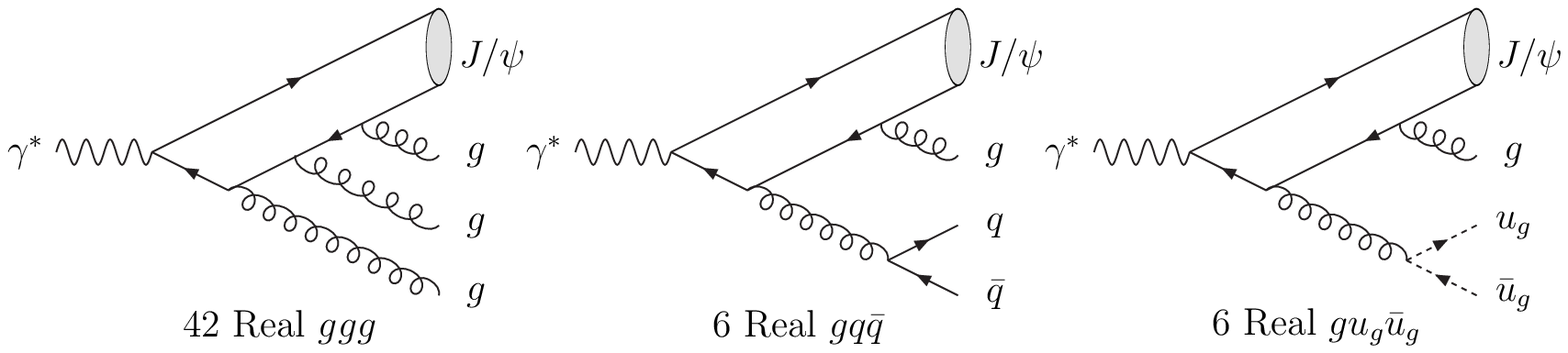}
\caption{\label{fig:real} Feynman diagrams  for the real correction
to $e^+ e^- \to J/\psi gg$. }
\end{figure}

We now present this calculation. We use {\tt
FeynArts}~\cite{feynarts} to generate Feynman diagrams and
amplitudes, {\tt FeynCalc}~\cite{Mertig:an} to handle amplitudes,
and {\tt LoopTools}~\cite{looptools} to evaluate the infrared-finite
scalar Passarino-Veltman integrals. Feynman diagrams for the Born,
virtual correction, and real correction are shown in
Fig.~\ref{fig:LO}, Fig.~\ref{fig:vir}, and Fig.~\ref{fig:real}. Note
$e^+ e^- \to J/\psi g c \bar{c}$ is excluded in the real correction,
because it should be included in the $J/\psi$ production associated
with open charm $e^+ e^- \to J/\psi + c \bar{c} + X$. Moreover, we
include ghost diagrams in the real correction because we choose
unphysical polarizations for the gluons in the final state.
There are generally ultraviolet(UV), infrared(IR), and Coulomb
singularities. Conventional Dimensional Regularization (CDR) with
$D=4-2\epsilon$ is adopted to regularize them.

The UV-divergences from self-energy and triangle diagrams are
removed by renormalization. The renormalization constants $Z_m$,
$Z_2$, and $Z_3$, which correspond  respectively to the charm quark
mass $m$, charm field $\psi_c$ and gluon field $A_\mu^a$, are
defined in the on-mass-shell(OS) scheme, while $Z_g$ corresponding
to the coupling $\alpha_s$ is defined in the
modified-minimal-subtraction($\overline{MS}$) scheme
\begin{eqnarray}
&&\hspace{-0.3cm}\delta Z_m^{OS}=-3C_F
\frac{\alpha_s}{4\pi}N_{\epsilon}\left[\frac{1}{\epsilon_{UV}}+\frac{4}{3}\right],\nonumber
\\ &&\hspace{-0.3cm}\delta Z_2^{OS}=-C_F
\frac{\alpha_s}{4\pi}N_{\epsilon}\left[\frac{1}
{\epsilon_{UV}}+\frac{2}{\epsilon_{IR}}+4\right],\nonumber\\
&&\hspace{-0.3cm}\delta Z_3^{OS}= \frac{\alpha_s}{4\pi}N_{\epsilon}
\left[(\beta_0(n_{lf})-2C_A)(\frac{1}{\epsilon_{UV}}-
\frac{1}{\epsilon_{IR}})-\frac{4T_f}{3\epsilon_{UV}}\right],\nonumber\\
&&\hspace{-0.3cm}\delta Z_g^{\overline{MS}}=-\frac{\beta_0(n_f)}{2}
\frac{\alpha_s}{4\pi}N_{\epsilon}\left[\frac{1}{\epsilon_{UV}}+\ln\frac{m^2}{\mu^2}\right],
\end{eqnarray}
where $N_{\epsilon}=\left(\frac{4\pi
\mu^2}{m^2}\right)^{\epsilon}\Gamma(1+\epsilon)$ is a overall factor
in our calculation,
 $\beta_0(n_f)=\frac{11}{3}C_A-\frac{4}{3}T_Fn_f$
is the one-loop coefficient of the QCD beta function, $n_f=4$ is the
number of active quark flavors, $n_{lf}=3$ is the number of light
quark flavors, and $\mu$ is the renormalization scale.

IR singularities coming from loop-integration and phase space
integration of real correction are found to cancel each other. We
use the method in \cite{Dittmaier:2003bc} to separate the soft and
collinear singularities in the virtual corrections, and then treat
the singular part analytically while the finite part numerically.

We use phase space slicing method\cite{Harris:2001sx} to extract
poles in the real correction. The method introduces a soft cut
$\delta_s$ and a hard collinear cut $\delta_c$ to the phase space.
Then the cut region can be partly integrated, and becomes some color
connected born cross sections multiplied by singular factors. While
the remaining region, hard non-collinear region, which is
non-singular, can be integrated using the standard Monte-Carlo
techniques. In order to make the method effective,
$\delta_c\ll\delta_s$ is needed as mentioned in\cite{Harris:2001sx}.
With a careful treatment for the two cuts, we verified that our
result is independent of the two cuts in a large range.

We then find that by considering all NLO virtual and real
corrections, and factoring the Coulomb singular term into the
$J/\psi$ wave function, we get an UV and IR finite cross section for
$e^+e^-\to J/\psi + gg$.

In the numerical calculation we use $\sqrt{s}=10.6{\rm GeV}$,
$\Lambda_{\overline{MS}}^{(4)}=338{\rm MeV}$, and $J/\psi$ wave
function squared at the origin $|R_{J/\psi}(0)|^2=1.01{\rm GeV}^3$,
which is extracted from the $J/\psi$ leptonic
width\cite{Amsler:2008zz} at NLO in $\alpha_s$: $
|R_{J/\psi}(0)|^2=\frac{9M^2_{J/\psi}}{16\alpha^2[1-(16/3)\alpha_s/
\pi]}\Gamma_{J/\psi\to e^+e^-} $.

\begin{figure}
\includegraphics[width=8.5cm]{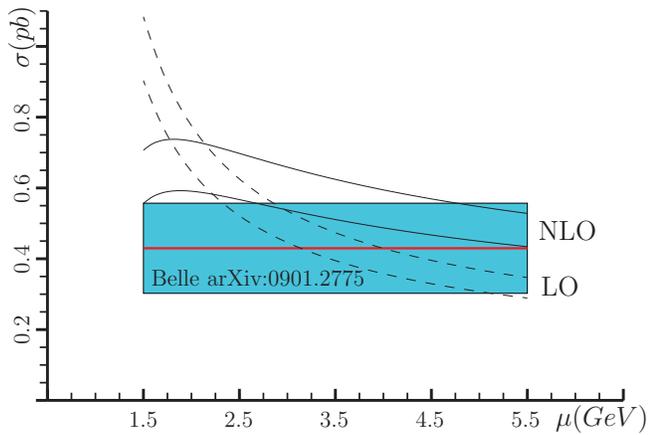}
\caption{\label{fig:psiggdepmu} Prompt cross sections of  $e^+ e^-
\to J/\psi g g$ as functions of the renormalization scale $\mu$  at
LO and NLO in $\alpha_s$. The upper curves correspond to $m=1.4~
{\rm GeV}$, and the lower ones correspond to $m=1.5~ {\rm GeV}$. }
\end{figure}

Taking $M_{J/\psi}=2m$ (at LO in $v$) and $m=1.4 ~{\rm GeV}$, we get
$\alpha_{s}(\mu)=0.267$ for $\mu=2m$, and the cross section at NLO
in $\alpha_s$ is $\sigma(e^+ e^- \to J/\psi g g)=0.496~{\rm pb}$,
which is a factor of 1.19 larger then the LO cross section $0.418
~{\rm pb}$.  If we set $\mu=\sqrt{s}/2$, then $\alpha_{s}=$0.211,
and the cross section is $0.394$~pb. Since the experimental data
correspond to the $J/\psi$ prompt production, in addition to the
direct production discussed above, we should also include the
feeddown contributions from higher charmonium states which decay
into $J/\psi$. Since for the P-wave states $\chi_{cJ}$ the direct
production rates in the non-$c\bar c$ associated process are
suppressed ($e^+e^-\to\gamma^*\to\chi_{cJ}gg$ are forbidden due to
charge parity conservation), and for the $\psi(nS)(n>2)$ states the
decay branching ratios into $J/\psi +X$ are negligible, we only need
to consider the $\psi(2S)$ feeddown contribution. This implies an
additional enhancing factor of $1.355$ should be multiplied
\cite{Zhang:2006ay}. In Fig.~\ref{fig:psiggdepmu} we show the prompt
production cross sections at LO and NLO as functions of the
renormalization scale $\mu$. We see that NLO QCD correction
substantially reduces the $\mu$ dependence,
and enhances the cross section by about 20\%.

\begin{figure}
\includegraphics[width=7.5cm]{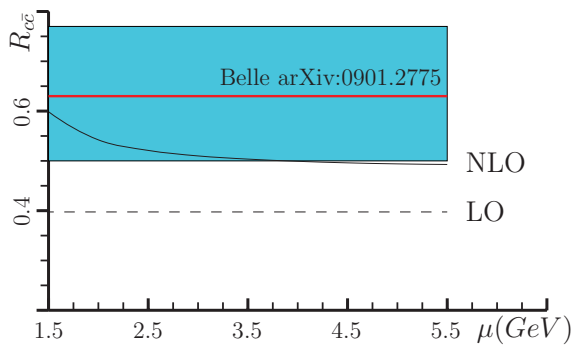}
\caption{\label{fig:rccdepmu} $R_{c \bar c}$ as a function of the
renormalization scale $\mu$  at LO and NLO in $\alpha_s$. Here
$m_c=1.4~$GeV.}
\end{figure}

The NLO cross sections $\sigma(e^+ e^- \to J/\psi c \bar c)$ were
calculated in Ref.~\cite{Zhang:2006ay}. We list the values of prompt
cross sections\cite{Zhang:2006ay} in TABLE \ref{table1}, together
with the prompt cross sections $\sigma(e^+ e^- \to J/\psi gg)$
obtained above.
Then we can get the ratio $R_{c\bar c}$ at LO and NLO in $\alpha_s$.
The dependence of $R_{c \bar c}$ on the renormalization scale $\mu$
is shown in Fig.~\ref{fig:rccdepmu}, where $m=1.4$~GeV is fixed. The
$\mu$ dependence for $\sigma(e^+ e^- \to J/\psi gg)$ is mild, while
for $\sigma(e^+ e^- \to J/\psi c \bar c)$ is strong. A reasonable
choice should be between $\mu=2m_c$ and $\mu=\sqrt{s}$/2, and more
preferably the latter.

\begin{table}[tb]
\begin {center}
\begin{tabular}{|c|c|c|c|c|}
 \hline
&$\mu=2.8$GeV& $\mu=2.8$GeV& $\mu=5.3$GeV&$\mu=5.3$GeV \\
 & LO & NLO & LO & NLO
 \\\hline
$\sigma(gg)$&0.57 & 0.67& 0.36& 0.53\\
$\sigma(c \bar c)$&0.38 & 0.71& 0.24&0.53 \\
$R_{c \bar c}$& 0.40& 0.51&0.40 & 0.50\\
\hline
\end{tabular}
\caption{Cross sections of prompt (feeddown included) $J/\psi gg$
(this Letter) and $J/\psi c\bar c$ (Re.\cite{Zhang:2006ay})
production in $e^+e^-$ annihilation at B factories in units of pb.}
 \label{table1}
\end {center}
\vspace{-0.5cm}
\end{table}

\begin{figure}
\includegraphics[width=7.5cm]{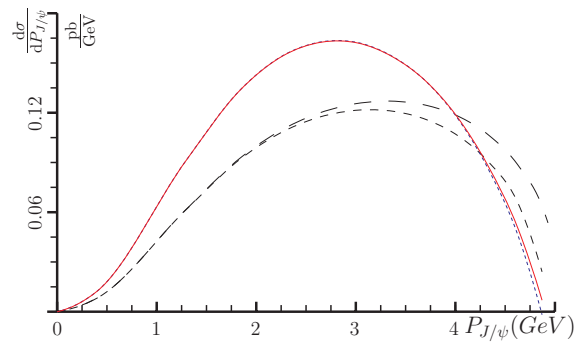}
\caption{\label{fig:ppsiDis}The cross section of  $e^+ e^- \to
J/\psi g g$ as functions of the $J/\psi$ momentum $P_{J/\psi}$. Here
$\mu=2.8~$GeV and $m=1.4~$GeV. The solid curve is the NLO+LL
prediction, and the dotted, short-dashed, and long-dashed curves are
the NLO, LO+LL, and LO predictions respectively.}
\end{figure}

Finally, we note that the large logarithms of $\log (1-z)$ appear at
the endpoint in NLO calculation, where
$z=E_{J/\psi}/E_{J/\psi}^{max}$. The leading logarithms (LL) have
been resumed in \cite{Lin:2004eu,Leibovich:2007vr}. Using a similar
approach, we define the differential cross section (and other
quantities) as \cite{Leibovich:2007vr}
\begin{eqnarray}\label{fulleq}
d \sigma_{\rm LO (NLO)+LL}&=& d \sigma_{\rm LO (NLO)}+ P[r,z]d
\sigma_{\rm resum} \nonumber
\\&&- P[r,z]\Big(d \sigma_{\rm resum} \Big)_{LO(NLO)},
\end{eqnarray}
where $(d \sigma_{\rm resum} )_{LO(NLO)}$ means expanding $d
\sigma_{\rm resum}$ in $\alpha_s$ to LO (NLO). To be consistent with
our previous calculation, we choose $\mu_c=\sqrt{2(1-z)}\mu_H$, and
$\mu_H=\mu=2m$. In Fig.~\ref{fig:ppsiDis} we show the cross sections
of $e^+ e^- \to J/\psi g g$ as functions of the $J/\psi$ momentum
$P_{J/\psi}$. The correction of LL resummation to the total cross
section is about $-6.6\%$ at LO and $0.5\%$ at NLO. But it becomes
large at the end point region when $z\to 1$, suppressing the LO
cross section and enhancing the NLO cross section. The LL
resummation changes the $J/\psi$ momentum distribution near the end
point, but has only a little effect on the total cross section. It
is interesting to note that with the NLO correction, the $J/\psi$
momentum spectrum becomes much softer than the LO result.


In summary, we find that by considering all NLO virtual  and real
corrections, and factoring the Coulomb singular term into the $c\bar
c$ bound state wave function, we get an ultraviolet, infrared, and
collinear finite cross section for the direct production of
$e^+e^-\to J/\psi + gg$ at $\sqrt{s}=10.6$~GeV, which enhances the
cross section by about 20\%.  Adding the feeddown contribution from
$\psi(2S)$ the prompt production cross section of $e^+e^-\to J/\psi
+ gg$ at NLO in $\alpha_s$ is found to be
$(0.67^{-0.13}_{+0.17})$~pb for $\mu=2m$ and
$(0.53^{-0.09}_{+0.12})$~pb for $\mu=\sqrt s/2$ with $m=(1.4\pm
0.1)$GeV. Together with the calculated $\sigma(e^+ e^- \to J/\psi c
\bar{c})$ at NLO in $\alpha_s$\cite{Zhang:2006ay}, we get $R_{c \bar
c}\approx 0.50$. The result significantly reduces the discrepancy
between theory and experiment. The effect of the leading logarithm
resummation near the end point on the $J/\psi+gg$ total cross
section is found to be small.

$Note$. Very recently Belle reported a new measurement with higher
statistics\cite{P. Pakhlov}:
\begin{eqnarray}
\sigma(e^+e^-\rightarrow J/\psi+c\bar
c)&\hspace{-0.2cm}=&\hspace{-0.2cm}(0.74\pm 0.08^{+0.09}_{-0.08})pb,
\\ \sigma(e^+e^-\rightarrow J/\psi+non(c\bar
c))&\hspace{-0.2cm}=&\hspace{-0.2cm}(0.43\pm 0.09\pm 0.09)pb.
\end{eqnarray}
The observed cross section of $e^+e^-\rightarrow J/\psi+non(c\bar
c)$ and $R_{c \bar c}$ are displayed in Fig.~\ref{fig:psiggdepmu}
and Fig.~\ref{fig:rccdepmu} with central values and error bands in
comparison with theoretical predictions. We see that, our
predictions (NLO with feeddown) for $\sigma(e^+e^-\rightarrow
J/\psi+gg)$ are consistent with the new measurement of
$\sigma(e^+e^-\rightarrow J/\psi+non(c\bar c))$ within certain
uncertainties. Moreover, the predicted $J/\psi$ momentum spectrum
also agrees with the experiment\cite{P. Pakhlov}. Importantly, our
result of $\sigma(e^+e^-\rightarrow J/\psi+non(c\bar c))$ indicates
that the calculated $\sigma(e^+e^-\rightarrow J/\psi+gg)$ has
already saturated the observed $\sigma(e^+e^-\rightarrow
J/\psi+non(c\bar c))$, hence leaving little room for the color-octet
contributions. These are also confirmed by a similar
study\cite{gong0901}, which agrees with ours.

~~~~~~~~~~~~~~~~~~
\begin{acknowledgments}
We thank C. Meng for helpful assistance and discussions in this
study, and G. Bodwin, B. Gong, J.W. Qiu, and J.X. Wang for useful
comments. This work was supported by the National Natural Science
Foundation of China (No 10675003, No 10721063, No 10805002).
\end{acknowledgments}


\end{document}